\documentclass[12pt,aps,prd,superscriptaddress,showpacs,longbibliography,floatfix,nofootinbib]{revtex4-1}

\usepackage[utf8]{inputenc}
\usepackage{slashed}
\pdfoutput=1

\usepackage{color}
\usepackage{graphicx}   
\usepackage{bm}
\usepackage{amsmath}
\usepackage{amsfonts}
\usepackage{eufrak}
\usepackage{hyperref}

\begin{document}

\title{Quantum Mechanical Out-Of-Time-Ordered-Correlators for the Anharmonic (Quartic) Oscillator}

\author{Paul Romatschke}
\affiliation{Department of Physics, University of Colorado, Boulder, Colorado 80309, USA}
\affiliation{Center for Theory of Quantum Matter, University of Colorado, Boulder, Colorado 80309, USA}

\begin{abstract}
  Out-of-time-ordered correlators (OTOCs) have been suggested as a means to study quantum chaotic behavior in various systems. In this work, I calculate OTOCs for the quantum mechanical anharmonic oscillator with quartic potential, which is classically integrable and has a Poisson-like energy-level distribution. For low temperature, OTOCs are periodic in time, similar to results for the harmonic oscillator and the particle in a box. For high temperature, OTOCs exhibit a rapid (but power-like) rise at early times, followed by saturation consistent with $2\langle x^2\rangle_T \langle p^2\rangle_T$ at late times. At high temperature, the spectral form factor decreases at early times, bounces back and then reaches a plateau with strong fluctuations.
\end{abstract}

\maketitle

\section{Introduction}

The framework to calculate OTOCs for quantum mechanical systems with general Hamiltonians has been set up in Ref.~\cite{Hashimoto:2017oit}. The simplest case is that of the harmonic oscillator, which can be treated analytically, and gives OTOCs that are purely oscillatory. More complicated examples that exhibit classical chaos such as the two-dimensional stadium billiard must be treated numerically and give OTOCs that are growing non-exponentially at early times followed by a saturation at late times \cite{Hashimoto:2017oit}.

Early-time exponential growth of OTOCs has been found in a system of non-linearly coupled oscillators \cite{Akutagawa:2020qbj}, which is expected to exhibit quantum chaos. More recently, however, it has been found that OTOCs show exponential growth in systems that do not possess quantum chaos, such as the inverted harmonic oscillator \cite{Bhattacharyya:2020art,Hashimoto:2020xfr}.

In the present work, I study OTOCs and spectral form factors for the case of the one-dimensional anharmonic quantum oscillator, which has not been done before.

\section{Out-of-time-ordered correlators in quantum mechanics}

This section largely follows the setup outlined in Ref.~\cite{Hashimoto:2017oit} to calculate both microcanonical and thermal OTOCs in quantum mechanics, which I will be using in the following. Given a Hamiltonian $H=H(\hat x,\hat p)$, the OTOC is defined as
\begin{equation}
  O_T(t)\equiv-\langle[x(t),p(0)]^2\rangle_T\,,
\end{equation}
where the subscript $T$ denotes calculation of the expectation value in a heat bath of temperature $T=\frac{1}{\beta}$. Specifically, introducing the microcanonical OTOC $c_n(t)$ as the OTOC for a fixed energy eigenstate $|n\rangle$ as in \cite{Hashimoto:2017oit}
\begin{equation}
  c_n(t)=-\langle n| [x(t),p(0)]^2 |n\rangle\,,
\end{equation}
this implies
\begin{equation}
  \label{otocdef}
  O_T(t)=Z^{-1}(\beta)\sum_{n=0}^\infty e^{-\beta E_n} c_n(t)\,,\quad Z(\beta)=\sum_{n=0}^\infty e^{-\beta E_n}\,,
\end{equation}
where $H|n\rangle=E_n|n\rangle$. The microcanonical OTOC may be expressed as 
\begin{equation}
  c_n(t)=\sum_{m=0}^\infty |b_{nm}(t)|^2\,,\quad b_{nm}(t)\equiv -i \langle n| [x(t),p]|m\rangle\,,
\end{equation}
and if the Hamiltonian is of the form
\begin{equation}
  \label{hamiltonian}
  H=p^2+V(x)\,,
\end{equation}
it was shown in Ref.~\cite{Hashimoto:2017oit} that this leads to
\begin{equation}
  b_{nm}(t)=\frac{1}{2}\sum_{k=0}^\infty x_{nk}x_{km}\left[\left(E_k-E_m\right) e^{-i t \left(E_k-E_n\right)}+\left(E_k-E_n\right)e^{i t \left(E_k-E_m\right)}\right]\,,\quad x_{nm}\equiv \langle n|x|m\rangle\,.
\end{equation}
As a consequence, knowledge of the energy eigenvalues $E_n$ and the matrix elements $x_{nm}$ is sufficient to calculate the quantum mechanical OTOC.

\subsection*{Spectral form factor in quantum mechanics}

Another quantity that is related to OTOCs is the spectral form factor.
Following \cite{Dyer:2016pou}, it that has been proposed as a handle on information loss and I will employ its normalized version defined as
\begin{equation}
  \label{ssf}
  g(\beta,t)\equiv \frac{|Z(\beta+i t)|^2}{Z^2(\beta)}\,,
  \end{equation}
where $Z(c)$ for complex $c$ is the analytically continued partition function defined in Eq.~(\ref{otocdef}). The normalization ensures that $g(\beta,0)=1$.

\section{Discrete ``solutions'' for the quartic  oscillator}
\label{discrete}

Let me consider a Hamiltonian such as (\ref{hamiltonian}) with an anharmonic oscillator potential $V(x)$ given by
\begin{equation}
  \label{eq:apot}
  V(x)=\frac{x^{2 N}}{4}\,,
\end{equation}
with $N\geq 1$ not necessarily integer. Widely known examples of this potential are the cases $N=1$ (the harmonic oscillator), $N=2$ (the quartic oscillator), $N=3$ (the sextic oscillator) and $N=\infty$ (particle in a box). Only $N=1$ and $N=\infty$ have known analytic solutions while for generic $N$ analytic solutions for the energy spectrum of the anharmonic quantum oscillator remain unknown. In the following, I concentrate on the case $N=2$ (the quartic oscillator) for which the Schr\"odinger equation becomes
\begin{equation}
  \label{Schro}
  \psi_n^{\prime\prime}(x)-\frac{x^4}{4}\psi_n(x)=-E_n \psi_n(x)\,,
\end{equation}
with $\psi_n(x)=\langle n |\psi\rangle$.

Now consider the auxiliary problem
\begin{equation}
  \label{aux}
  \chi^{\prime\prime}(x)-\frac{x^4}{4}\chi(x)=-\gamma x \chi(x)\,,\quad x\geq 0
\end{equation}
which has a discrete solution spectrum that is spanned by the associated generalized Laguerre polynomials. Specifically, separating wave-functions into parity-even (+) and parity-odd (-) solutions, (\ref{aux}) is solved by
\begin{equation}
  \chi^+(x)=e^{-\frac{x^3}{6}}L_i^{\left(-\frac{1}{3}\right)}\left(\frac{x^3}{3}\right)\,,\quad
  \chi^-(x)=x e^{-\frac{x^3}{6}}L_i^{\left(+\frac{1}{3}\right)}\left(\frac{x^3}{3}\right)\,,
\end{equation}
where $i=0,1,2,\ldots$ and the respective eigenvalues $\gamma$ in (\ref{aux}) take the values $\gamma^+_i=1+3i$, $\gamma^-_i=2+3i$. Expanding wave-functions $\psi_n(x)$ of the original problem (\ref{Schro}) in terms of the auxiliary functions  thus leads to
\begin{equation}
  \label{ansatz}
  \psi_{2n}(x)=e^{-\frac{x^3}{6}}  \sum_{i=0}^K c_i ^{(2n)} L_i^{\left(-\frac{1}{3}\right)}\left(\frac{x^3}{3}\right)\,,\quad
  \psi_{2n+1}(x)=x e^{-\frac{x^3}{6}} \sum_{i=0}^K c_i ^{(2n+1)} L_i^{\left(+\frac{1}{3}\right)}\left(\frac{x^3}{3}\right)\,,
\end{equation}
with $K\rightarrow \infty$ and obvious modifications for $x<0$. I will refer to (\ref{ansatz}) as the ``Laguerre-transform'' of $\psi(x)$, and to $c_i$ as the ``Laguerre-coefficients''. Plugging (\ref{ansatz}) back into (\ref{Schro}), the Laguerre coefficients $c_i^{(n)}$ have to fulfill
\begin{eqnarray}
  \label{mf}
  E_{2n}\sum_{i=0}^K c_i ^{(2n)} L_i^{\left(-\frac{1}{3}\right)}\left(\frac{x^3}{3}\right) &=& x \sum_{i=0}^K (1+3i) c_i ^{(2n)} L_i^{\left(-\frac{1}{3}\right)}\left(\frac{x^3}{3}\right)\,,\\
  E_{2n+1}\sum_{i=0}^K c_i ^{(2n+1)} L_i^{\left(\frac{1}{3}\right)}\left(\frac{x^3}{3}\right) &=& x \sum_{i=0}^K (2+3i) c_i ^{(2n+1)} L_i^{\left(\frac{1}{3}\right)}\left(\frac{x^3}{3}\right)\,.
\end{eqnarray}

After using the steps detailed in appendix \ref{details}, the energy eigenvalues $E_{n}$ and Laguerre coefficients $c_i$ are found to be related to the eigenvalues and eigenvectors of $A_{ij},B_{ij}$ given in Eqns.~(\ref{ABmatrices}):
\begin{eqnarray}
  \label{evproblem}
  A_{ij}\tilde c_j^{(2n)}=\frac{3^{\frac{1}{3}}}{\Gamma\left(\frac{1}{3}\right)E_{2n}} \tilde c_i^{(2n)}\,,
  \quad
  B_{ij}\tilde  c_j^{(2n+1)}=\frac{3^{\frac{1}{3}}}{E_{2n+1}} \tilde c_i^{(2n+1)}\,,
\end{eqnarray}
where $i,j=0,1,2,\ldots K$ and I have rescaled the Laguerre coefficients as
\begin{equation}
  \label{rescaling}
  c_i^{(2n)}=\tilde{c}_i^{(2n)} \sqrt{\frac{i!}{(1+3i)\Gamma\left(\frac{2}{3}+i\right)}}\,,\quad
  c_i^{(2n+1)}=\tilde{c}_i^{(2n+1)} \sqrt{\frac{i!}{(2+3i)\Gamma\left(\frac{4}{3}+i\right)}}\,.
  \end{equation}

A full solution for the quartic oscillator requires finding the eigenvalues and eigenvectors for $K\rightarrow \infty$, for which $A,B$ become infinite matrices. Keeping $K$ finite can either be viewed as an approximation to the full solution, or, to a solution at a discrete set of points. Specifically, solving the eigenproblem (\ref{evproblem}) for fixed and finite K, leads to K parity-even and K parity-odd wave-functions $\tilde \psi_{n}(x)$ that each are exact solutions to (\ref{Schro}) at 2K points $x=\pm x_{1},\pm x_{2}\,\ldots, \pm x_{K}$. In many respects, this is reminiscent of constructing continuum solutions out of Fourier series.

For pedagogical reasons, some of the above properties are discussed for analytically tractable case K=1 in the next section.

\section{Analytically tractable approximation for $c_0(t)$}

For K=1, the eigenvalue problems (\ref{evproblem}) is analytically solvable. The non-vanishing matrix elements (\ref{matel}) are evaluated to be
\begin{eqnarray}
  \label{k2matel}
  x_{01}=\simeq 0.81\,,\quad
  x_{03}=\simeq -0.01\,,\quad
  x_{21}=\simeq 0.87\,,\quad
  x_{23}=\simeq 1.65\,,
\end{eqnarray}
cf. appendix \ref{details} for details. Note that $x_{03}$ is an order of magnitude smaller than the other matrix elements, such that neglecting contributions involving $x_{03}$ is a good approximation. As a consequence, to good approximation,
\begin{eqnarray}
  b_{00}(t)&\simeq&x_{01}^2 (E_1-E_0) \cos{(E_1-E_0)t}\,,\nonumber\\
  b_{02}(t)&\simeq&\frac{x_{01}x_{21}}{2}\left[(E_1-E_2)e^{-i t (E_1-E_0)}+(E_1-E_0)e^{i t (E_1-E_2)}\right]\,,
\end{eqnarray}
such that
\begin{eqnarray}
  \label{anac0}
  c_0(t)&\simeq& \frac{x_{01}^4(E_1-E_0)^2}{2}+\frac{x_{01}^2x_{21}^2}{4}\left[(E_1-E_0)^2+(E_2-E_1)^2\right]\\
 && +\frac{x_{01}^4(E_1-E_0)^2}{2} \cos{2 t (E_1-E_0)}
  -\frac{x_{01}^2x_{21}^2}{2} (E_1-E_0)(E_2-E_1)\cos{t(2 E_1-E_0-E_2)}\,.\nonumber
\end{eqnarray}
This results suggests that $c_0(t)$ is the superposition of two harmonics  plus a constant. I find that (\ref{anac0}) is  correct semi-quantitatively also when $K\rightarrow \infty$, when replacing the numerical values $E_0,E_1,E_2,x_{01},x_{21}$ accordingly from $K=1$ to higher K.

\section{Numerical Results}

\begin{figure*}[t]
  \includegraphics[width=\linewidth]{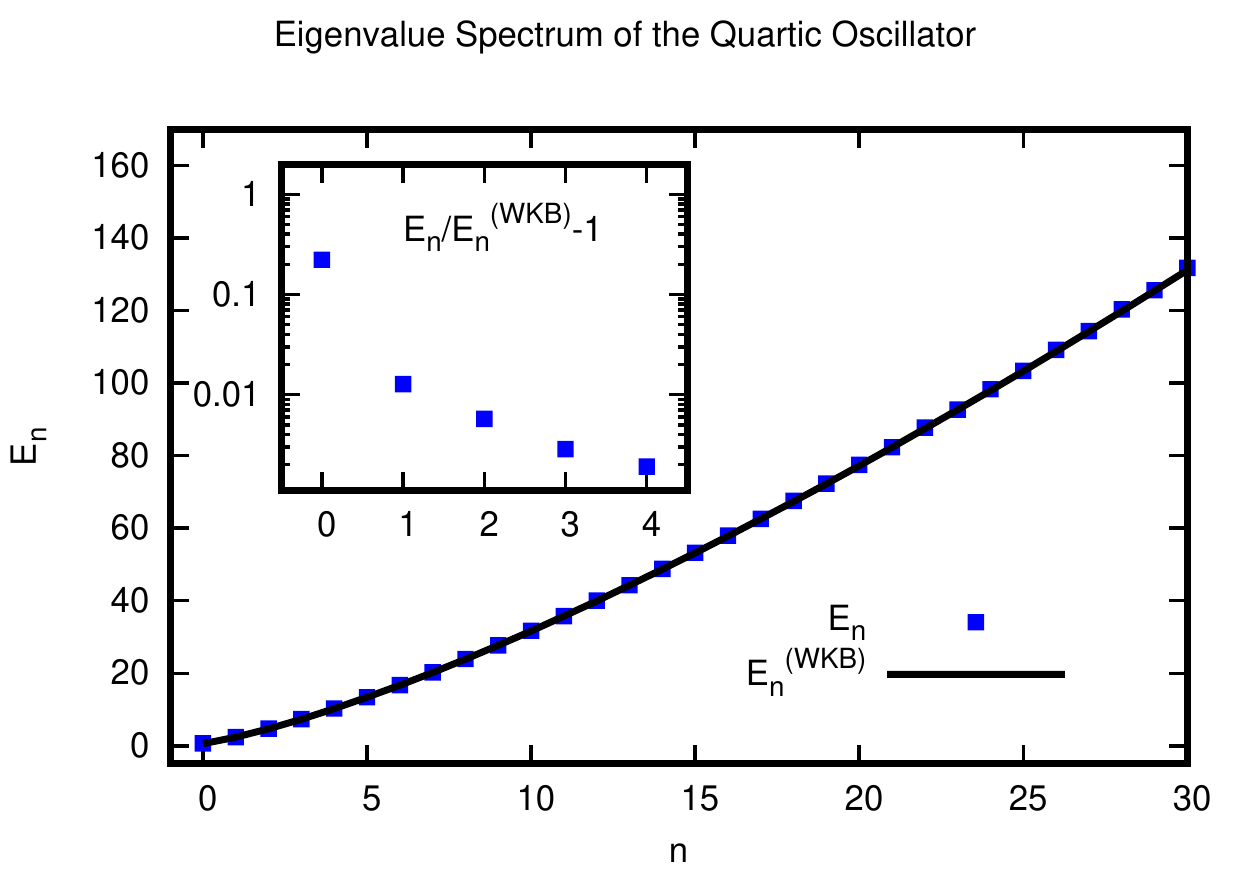}
  \caption{\label{fig1} Eigenvalue spectrum $E_n$ of the anharmonic quartic oscillator, compared to the WKB approximation as a function of $n$. Inset shows $E_n/E_n^{\rm (WKB)}-1$ to demonstrate that eigenvalues rapidly converge to the WKB values (note the logarithmic scale).}
\end{figure*}

Except for the ground state where reasonable approximations can be found\footnote{See e.g. Appendix \ref{main}.}, Eq.~(\ref{evproblem}) is hard to solve analytically n the limit $K\rightarrow \infty$. However, for generic K, (\ref{evproblem}) is amenable to efficient numerical solution using readily available eigenvalue packages. In practice, I use $K=4096$ and vary by a factor of two to test for numerical sensitivity w.r.t. finite K, showing only results that do not exhibit such sensitivity to the naked eye. As an example, results for the eigenvalue spectrum of the quartic oscillator are shown in Fig.~\ref{fig1}. For comparison, results from the WKB approximation \cite{Bender:1977dr,Hioe:1978jj} are shown in Fig.~\ref{fig1}, which in my units becomes
\begin{equation}
  \label{ewkb}
  E_n^{\rm (WKB)}=\frac{3 \pi^2 \left[3\left(n+\frac{1}{2}\right)^4\right]^{\frac{1}{3}}}{\Gamma\left(\frac{1}{4}\right)^{\frac{8}{3}}}\,.
\end{equation}
It is a well-known feature that the eigenvalue spectrum of the quartic oscillator is numerically close to the WKB spectrum for all but lowest-lying eigenvalues (see inset in Fig.~\ref{fig1}). However, the same is not true for the eigenfunctions $\psi_n(x)$ which in the WKB approximation feature a singularity close to the classical turning point.

\begin{figure*}[t]
  \includegraphics[width=\linewidth]{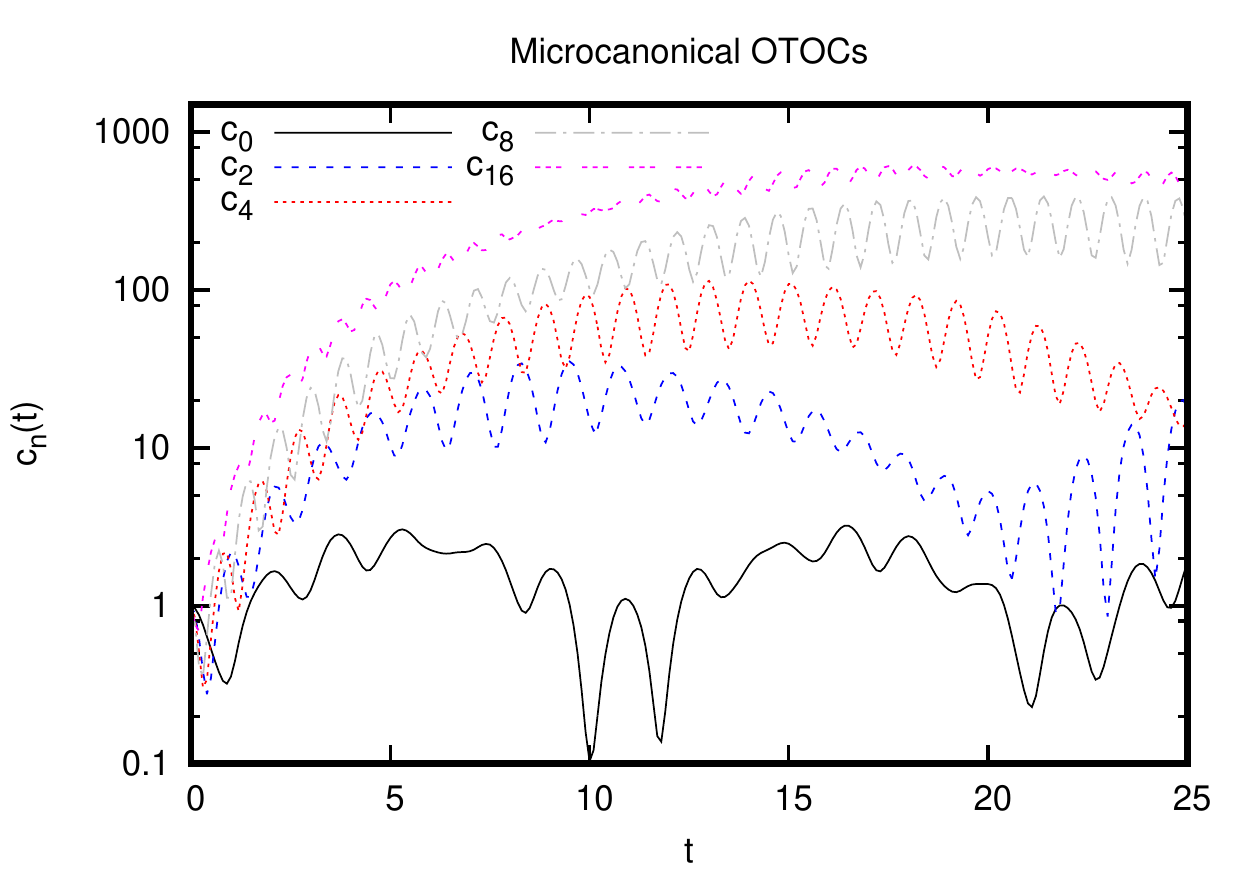}
  \caption{\label{fig2} Microcanonical OTOCs $c_n(t)$ as a function of time for $n=0,2,4,8,16$ (OTOCs for odd $n$ are qualitatively similar). }
\end{figure*}

Using the numerically determined Laguerre coefficients $\tilde c_i$, I find that the matrix elements (\ref{matel}) for generic K are strongly peaked for nearest energy-levels only, e.g.
\begin{equation}
  x_{2n,2m+1}\simeq \#_1\delta_{n,m}+\#_2\delta_{n,m+1}\,,
\end{equation}
with $\#_1,\#_2$ two numbers of order unity. This is similar to what was found in (\ref{k2matel}) for K=1. Therefore, I find for $n\geq 2$
\begin{equation}
  c_{n}(t)\simeq |b_{n,n-2}(t)|^2+|b_{n,n}(t)|^2+|b_{n+2,n}(t)|^2\,,
\end{equation}
and $b_{nm}(t)$ similarly dominated by nearest-neighbor energy levels. As a consequence, evaluation of $c_n(t)$ is dominated by matrix elements $x_{ij}$ with $i,j\simeq n$, which greatly reduces the computational cost whenever K becomes large. In practice, I determine and use all matrix elements up to $x_{64,65}$ when calculating OTOCs, resulting in a very good approximation for $c_n(t)$ with $n\leq 32$.


\begin{figure*}[t]
  \includegraphics[width=\linewidth]{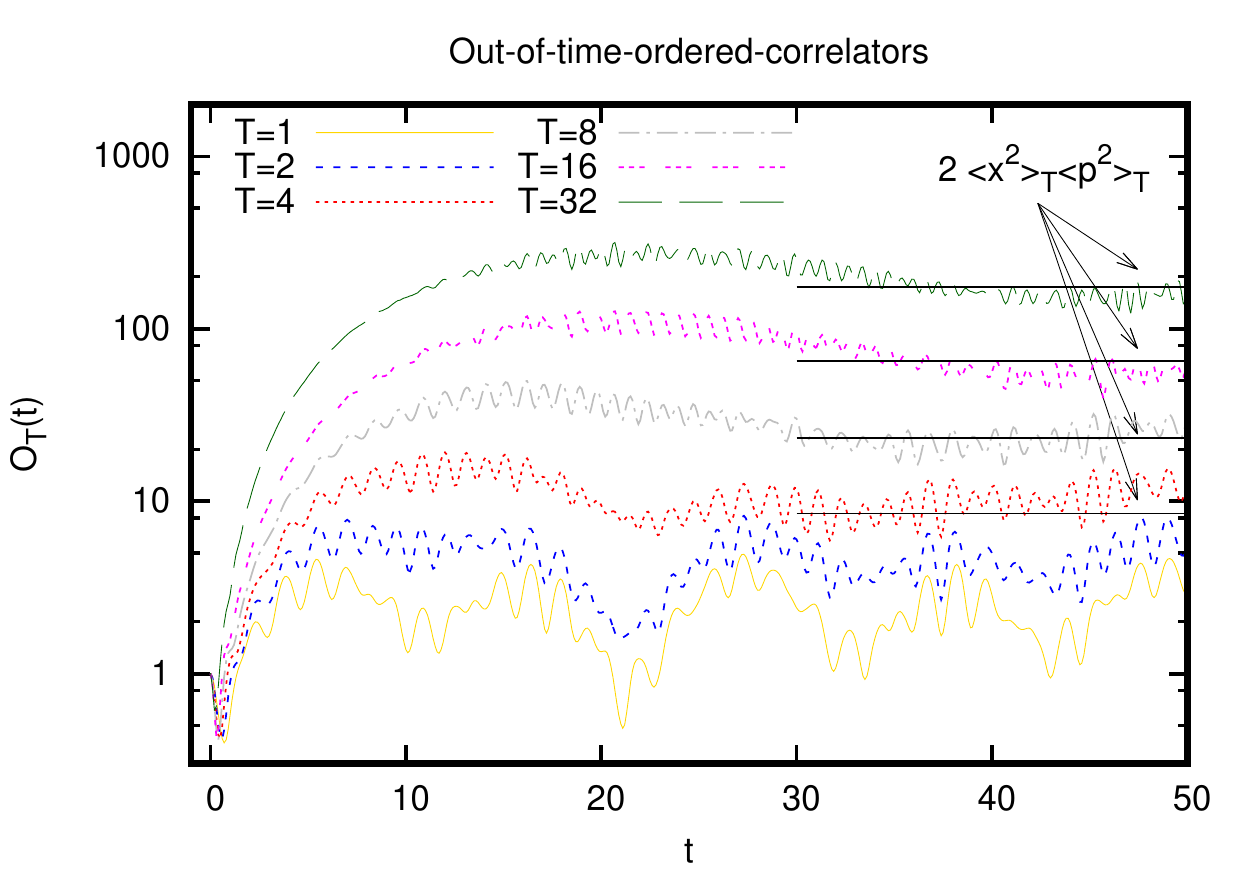}
  \caption{\label{fig3} OTOCs $O_T(t)$ as a function of time for various temperatures $T$. For higher temperatures, there is a rapid rise at early times followed by a plateau-like behavior at late times. Note that this rapid rise is not exponential, but rather power-law like. For comparison, the expected behavior of $O_T(\infty)$ for a quantum chaotic system (\ref{qcs}) is shown for $T=4,8,16,32$.}
\end{figure*}

Examples for low-lying microcanonical OTOCs $c_n(t)$ are shown in Fig.~\ref{fig2}. First note that $c_0(t)$ qualitatively agrees with the analytic result (\ref{anac0}) in that it behaves as the superposition of two harmonics plus a constant. For higher $n$, $c_n(t)$ exhibit a rapid rise at early times, followed by a qualitatively similar behavior: a constant plus two harmonics. However, results for $c_n(t)$ shown in Fig.~\ref{fig2} indicate that as $n$ increases, the period of the high-frequency harmonic does not change much, while the period of the low-frequency harmonic increases with $n$. This feature can be qualitatively understood when identifying the high-frequency harmonic with $\cos[2 (E_n-E_{n-1})t]$ and the low-frequency harmonic with $\cos[(2 E_{n}-E_{n-1}-E_{n+1})t]$, cf. Eq.~(\ref{anac0}), and using (\ref{ewkb}) to find
\begin{equation}
  \lim_{n\rightarrow \infty}  \left(E_{n}-E_{n-1}\right)\propto n^{\frac{1}{3}}\,,
  \quad
  \lim_{n\rightarrow \infty}  \left(E_{n-1}+E_{n+1}-2 E_n\right)\propto n^{-\frac{2}{3}}\,.
  \end{equation}

Since Fig.~\ref{fig2} also indicates that the amplitude of the high-frequency harmonic contributing to $c_n(t)$ is considerably less than those of the low-frequency parts for $n\gg 1$, the resulting behavior of $c_n(t)$ is that of a rapid rise at early times followed by a near-constant behavior at late times.

The resulting OTOC (\ref{otocdef}) for various temperatures is shown in Fig.~\ref{fig3}. Thermal averaging of the microcanonical $c_n(t)$ seems to have the effect of further reducing the amplitude of the harmonic contributions such that at temperatures $T\geq 8$, Fig.~\ref{fig3} suggests a rapid early-time rise followed by a plateau at late times for $O_T(t)$. While the early-time rise is clearly not a simple exponential, this qualitative behavior of $O_T(t)$ (rapid rise, saturation) has been associated with quantum chaotic behavior in systems that exhibit chaos, cf. Ref.~\cite{Akutagawa:2020qbj}.  However, for the case of the quartic oscillator, the system is classically integrable, and hence not expected to exhibit quantum chaos.

For quantum chaotic system, it is expected that at late times \cite{Akutagawa:2020qbj}
\begin{equation}
  \label{qcs}
\lim_{t\rightarrow \infty}  O_T(t)=2\langle x^2\rangle_T \langle p^2\rangle_T\,,
\end{equation}
where $\langle {\cal O}\rangle_T$ again denotes the thermal expectation value of the operator ${\cal O}$. Using the quantum virial theorem, it is straightforward to relate the expectation value of the momentum-operator squared to
\begin{equation}
  \langle n| p^2|n\rangle=\frac{2}{3}E_n\,,
\end{equation}
whereas the expectation value of the position operator is given by (\ref{eq:position}). The result from evaluating (\ref{qcs}) compared to $O_T(t)$ is shown for higher temperatures in Fig.~\ref{fig3}. This comparison indicates that the apparent saturation of $O_T(t)$ at high temperatures and late times is roughly consistent with (\ref{qcs}).

\begin{figure*}[t]
  \includegraphics[width=\linewidth]{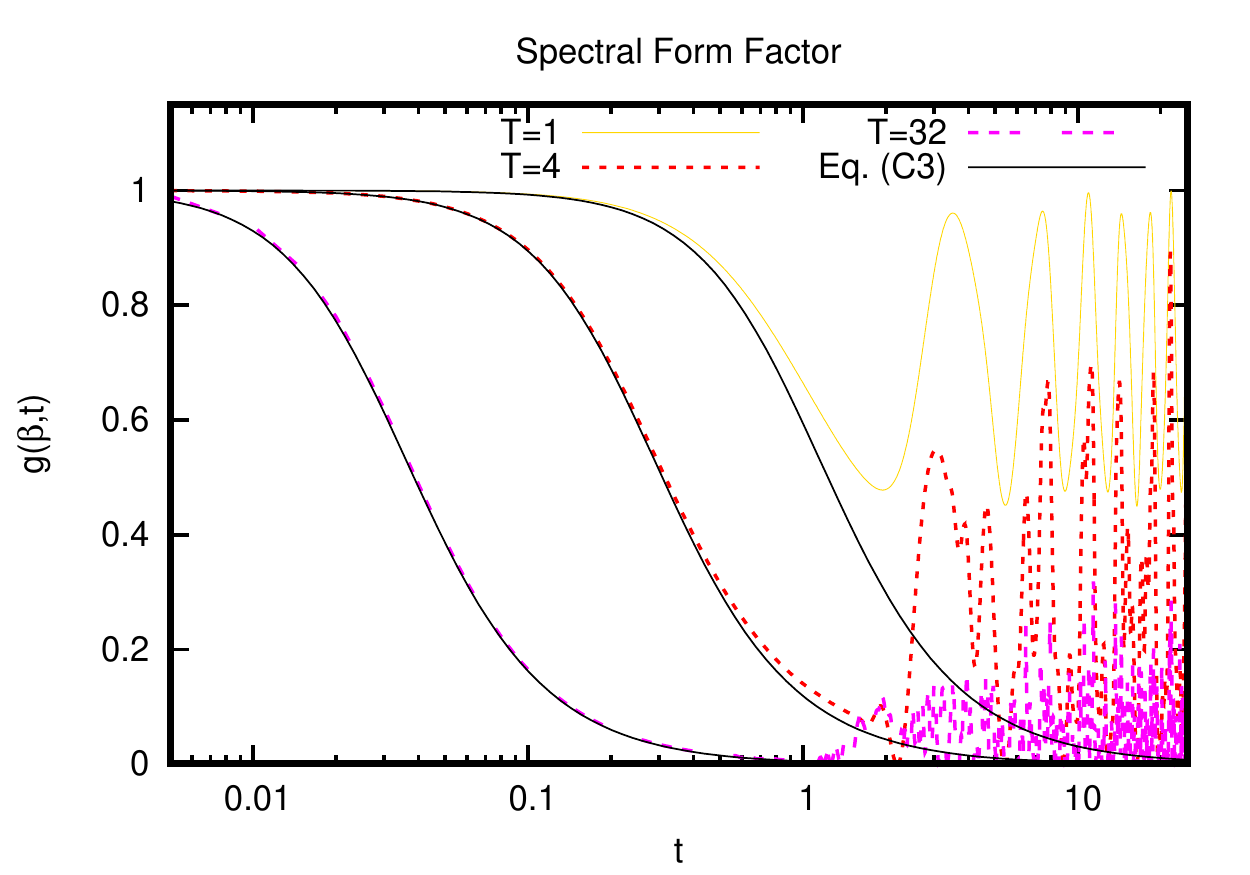}
  \caption{\label{fig4} Spectral form factor $g(\beta,t)$, normalized to $g(\beta,0)=1$ as a function of time for three different temperatures $T=1,4,32$. For high temperatures, the behavior of the spectral form factor is very much reminiscent of that from random matrices, cf. Ref.~\cite{Dyer:2016pou}. For comparison, the analytic high-temperature limit of the form factor (\ref{bsf}) is also shown.}
\end{figure*}

Finally, to complete the picture I also show the spectral form factor (\ref{ssf}) in Fig.~\ref{fig4}. At low temperatures, $g(\beta,t)$ seems to first decrease and then bounce back to its value at early times, while for high temperatures, $g(\beta,t)$ first drops, then rises again followed by a plateau with strong fluctuations. The high temperature behavior of $g(\beta,t)$ is not that dissimilar from what has been observed for a single random matrix, cf. Ref.~\cite{Dyer:2016pou}.

\begin{figure*}[t]
  \includegraphics[width=\linewidth]{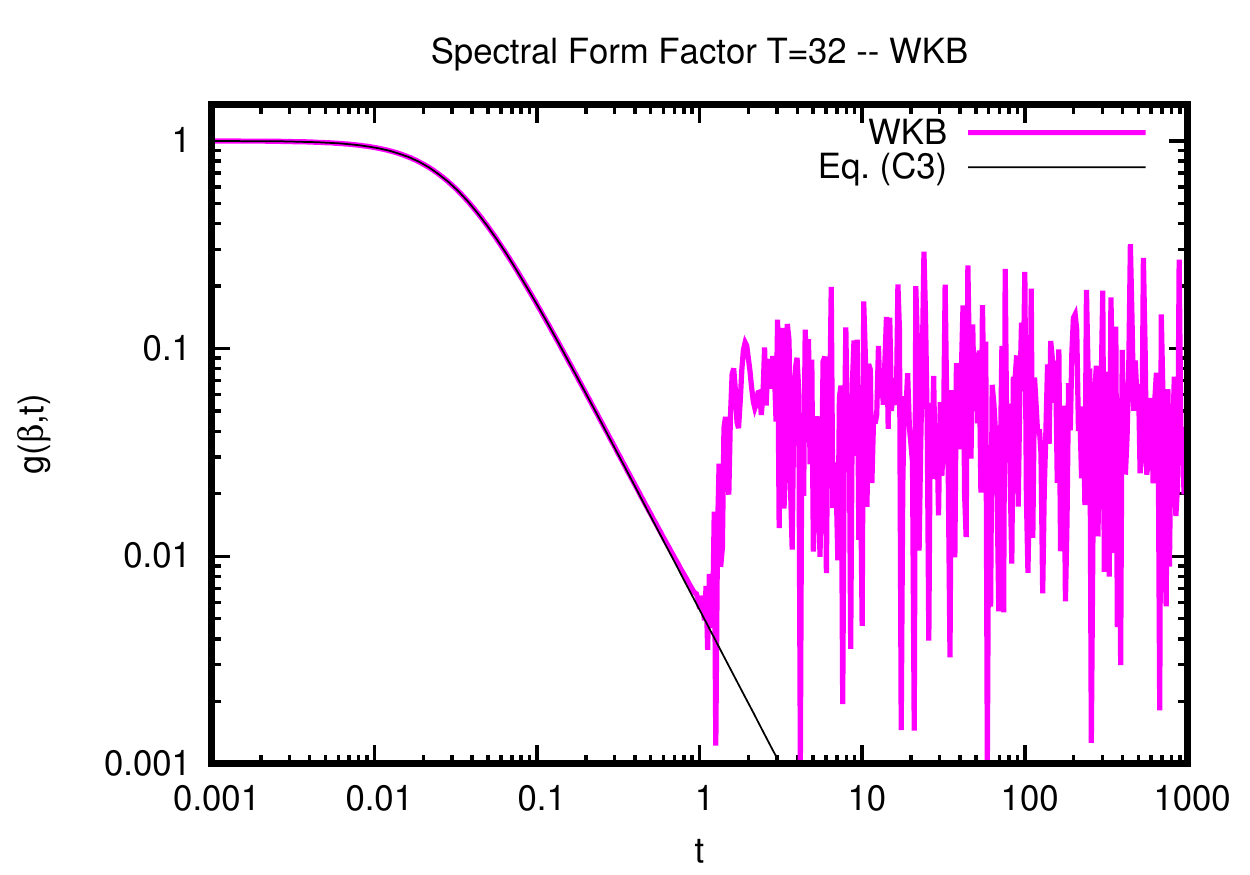}
  \caption{\label{fig5} Spectral form factor $g(\beta,t)$, normalized to $g(\beta,0)=1$ using the WKB energy eigenvalues (\ref{ewkb}) for $T=32$.  For comparison, the analytic high-temperature limit of the form factor (\ref{bsf}) is also shown.}
\end{figure*}

The behavior for small times can be calculated in a straightforward manner from the exact partition function in the high temperature limit (\ref{bsf}).  As can be seen from Fig.~\ref{fig4}, the analytic result for the spectral form factor matches the numerical results at early times, but deviates at late times. To keep the non-trivial late-time dependence in the high-temperature approximation, one can consider evaluating $g(\beta,t)$ using the WKB energy eigenvalues (\ref{ewkb}) in (\ref{otocdef}). The result for $T=32$ is shown in Fig.~\ref{fig5}. From this figure, it can be seen that the WKB form factor matches the analytic result (\ref{bsf}), while faithfully reproducing the late-time ``bounce'' and fluctuating plateau at late times for high temperature.

\section{Summary and Conclusions}

In this work, I studied the quantum-mechanical out-of-time-ordered-correlators for quartic interaction potential. It was found that at low temperature, OTOCs are periodic, while at high temperature OTOCs exhibit rapid power-law growth followed by an apparent saturation consistent with the temperature-dependent value $2 \langle x^2\rangle_T \langle p^2\rangle_T$. At high temperatures, the spectral form factor for this theory decreases at early times, followed by a bounce and a plateau with strong fluctuations. It was found that the early-time decrease can be understood from the analytic high-temperature limit of the partition function.

In conclusion, many interesting observables for one-dimensional quantum mechanics with anharmonic oscillator potential can be readily calculated using a spectral (Laguerre) decomposition. This may have implications for our understanding of quantum chaos, in particular when paired with the search for theories with gravitational duals.

While the present study was performed for quartic oscillator potential, all of the steps presented here can be repeated for potentials $V(x)\propto x^N$ with arbitrary (even non-integer) N, cf. Eq.~(\ref{eq:apot}). Since it may be interesting in the future to study properties such as the power-law rise in early-time OTOCs or the dip-time in the spectral form factor as a function of the potential index $N$, I made the numerical codes used to generate the plots in this study publicly available at \cite{codedown}.

  \section*{Acknowledgments}

  This work was supported by the Department of Energy, DOE award No DE-SC0017905. I am indebted to Koji Hashimoto for helping me put this work into context and encouraging me to make it publicly available.

  \begin{appendix}
    \section{Some details on solving the eigenvalue problem}
    \label{details}

    This appendix collects some of the more technical details of solving the eigenvalue problem discussed in section \ref{discrete}, starting with equations (\ref{mf}. Multiplying these equations with $e^{-\frac{x^3}{3}} L_j^{\left(-\frac{1}{3}\right)}$, $e^{-\frac{x^3}{3}} x^2 L_j^{\left(\frac{1}{3}\right)}$, respectively, changing variables to $z=\frac{x^3}{3}$ and integrating using \cite{Carlitz}
\begin{eqnarray}
  \label{carlitz}
  \int_0^\infty dz e^{-z} z^{\kappa}L_i^{(\kappa)}(z)L_{j}^{(\kappa)}(z)&=&\frac{\Gamma(i+\kappa+1)}{i!} \delta_{ij}\,,\\
  \int_0^\infty dz e^{-z} z^{\kappa}L_i^{(\phi)}(z)L_{j}^{(\rho)}(z)&=&\Gamma\left(1\!+\!\kappa\right) \frac{\left(\phi\!-\!\kappa\right)_i \left(\rho\!-\!\kappa\right)_j}{i! j!} \left._3 F_2\right. \left(\begin{array}{ccc}
    -i & -j & 1\!+\!\kappa\\
    &    1\!+\!\kappa\!-\!\phi\!-i & 1\!+\!\kappa\!-\!\rho\!-\!j
  \end{array}
  \right)\,,\nonumber
\end{eqnarray}
where $(a)_i$ denotes the Pochhammer symbol and $\left._3F_2\right.$ is a generalized hypergeometric function evaluated at unit argument, leads to 
\begin{eqnarray}
  \frac{\Gamma\left(\frac{2}{3}+j\right) (1+3j) 3^{\frac{1}{3}}}{j! E_{2n} \Gamma\left(\frac{1}{3}\right)}c_j^{(2n)}=\sum_{i=0}^K \frac{\left(\frac{1}{3}\right)_i \left(\frac{1}{3}\right)_j}{i! j!} \left._3F_2\right.\left(
  \begin{array}{ccc}
    -i & -j & \frac{1}{3}\\
    & \frac{2}{3}-i & \frac{2}{3}-j
    \end{array}
  \right) c_i^{(2n)}\,,\nonumber\\
  \frac{\Gamma\left(\frac{4}{3}+j\right) (2+3j) 3^{\frac{1}{3}}}{j! E_{2n+1} }c_j^{(2n+1)}=\sum_{i=0}^K \frac{\left(\frac{1}{3}\right)_i \left(\frac{1}{3}\right)_j}{i! j!} \left._3F_2\right.\left(
  \begin{array}{ccc}
    -i & -j & 1\\
    & \frac{2}{3}-i & \frac{2}{3}-j
    \end{array}
  \right) c_i^{(2n+1)}\,.
\end{eqnarray}
Writing
\begin{eqnarray}
  \label{ABmatrices}
  A_{ij}=\frac{\left(\frac{1}{3}\right)_i\left(\frac{1}{3}\right)_j}{\sqrt{i! j! (1+3i)(1+3j)\Gamma\left(\frac{2}{3}+i\right)\Gamma\left(\frac{2}{3}+j\right)}}
  \left._3F_2\right.\left(
  \begin{array}{ccc}
    -i & -j & \frac{1}{3}\\
    & \frac{2}{3}-i & \frac{2}{3}-j
    \end{array}
  \right)\,,\nonumber\\
   B_{ij}=\frac{\left(\frac{1}{3}\right)_i\left(\frac{1}{3}\right)_j}{\sqrt{i! j! (2+3i)(2+3j)\Gamma\left(\frac{4}{3}+i\right)\Gamma\left(\frac{4}{3}+j\right)}}
  \left._3F_2\right.\left(
  \begin{array}{ccc}
    -i & -j & 1\\
    & \frac{2}{3}-i & \frac{2}{3}-j
    \end{array}
  \right)\,,
\end{eqnarray}
one obtains Eq.~(\ref{evproblem}) in the main text. Wavefunctions are suitably normalized if $\int_{-\infty}^\infty \psi_n^2(x)=1$, which leads to the following normalization conditions for $\tilde c_i$:
\begin{eqnarray}
  \label{cnorm}
  1=2 \int_0^\infty dx \psi_{2n}^2(x)=2\times 3^{-\frac{2}{3}}\Gamma\left(\frac{1}{3}\right)\sum_{ij}\tilde c_i^{(2n)} A_{ij} \tilde c_j^{(2n)}=\frac{2\times 3^{-\frac{1}{3}}}{E_{2n}}\sum_{i=0}^K \tilde c_i^{(2n)}\tilde c_i^{(2n)}\,,\\
  1=2 \int_0^\infty dx \psi_{2n+1}^2(x)=2\sum_{ij}\tilde c_i^{(2n+1)} B_{ij} \tilde c_j^{(2n+1)}=\frac{2\times 3^{\frac{1}{3}}}{E_{2n+1}}\sum_{i=0}^K \tilde c_i^{(2n+1)}\tilde c_i^{(2n+1)}\,.
  \end{eqnarray}

I close this section by pointing out that because of parity-symmetry, matrix elements $x_{nm}$ require $n$ to be even and $m$ to be odd (or vice-versa). Without loss of generality assuming $n$ to be even, one finds using (\ref{carlitz}), \cite[Eq.~16.4.3]{NIST:DLMF}
\begin{eqnarray}
  \label{matel}
  x_{nm}&=&2 \sum_{ij} c_i^{(n)}c_j^{(m)} \int_0^\infty dx e^{-\frac{x^3}{3}}  L_i^{\left(-\frac{1}{3}\right)}\left(\frac{x^3}{3}\right)L_j^{\left(+\frac{1}{3}\right)}\left(\frac{x^3}{3}\right)\,x^2\,,\nonumber\\
  &=&2 \sum_{ij} c_i^{(n)}c_j^{(m)} \frac{\left(-\frac{1}{3}\right)_i \left(\frac{1}{3}\right)_j}{i! j!} \left._3F_2\right.\left(
  \begin{array}{ccc}
    -i & -j & 1\\
&    \frac{4}{3}-i & \frac{2}{3}-j
  \end{array}\right)\,,\nonumber\\
&=&  2 \sum_{ij} \tilde c_i^{(n)}\tilde c_j^{(m)} \frac{\left(-\frac{1}{3}\right)_i \left(\frac{1}{3}\right)_j}{\sqrt{i! j! (1+3i)(2+3j)\Gamma\left(\frac{2}{3}+i\right)\Gamma\left(\frac{4}{3}+j\right)}}\frac{(1-3i)(1+3j)}{1-3i+3j}\,.
  \end{eqnarray}

Other matrix elements that will be needed in the following are
\begin{eqnarray}
  \label{eq:position}
  &\langle 2n|x^2|2n\rangle=\hspace*{10cm}&\nonumber\\
  &\sum_{ij}\frac{\tilde c_i^{(2n)}\tilde c_j^{(2n)}}{|\tilde c^{(2n)}|^2}\frac{E_{2n} 3^{\frac{1}{3}}\times \left(-\frac{1}{3}\right)_i\left(-\frac{1}{3}\right)_j}{\sqrt{i! j! (1+3i)(1+3j)\Gamma\left(\frac{2}{3}+i\right)\Gamma\left(\frac{2}{3}+j\right)}} \left._3F_2\right.
    \left(\begin{array}{ccc}
      -i & -j & 1\\
      &\frac{4}{3}-i&\frac{4}{3}-j
    \end{array}\right)\,,&\nonumber\\
    &\langle 2n+1|x^2|2n+1\rangle=\hspace*{10cm}&\nonumber\\
    &\sum_{ij}\frac{\tilde c_i^{(2n+1)}\tilde c_j^{(2n+1)}}{|\tilde c^{(2n+1)}|^2}\frac{E_{2n+1} 3^{\frac{1}{3}}\Gamma\left(\frac{5}{3}\right)\times \left(-\frac{1}{3}\right)_i\left(-\frac{1}{3}\right)_j}{\sqrt{i! j! (2+3i)(2+3j)\Gamma\left(\frac{4}{3}+i\right)\Gamma\left(\frac{4}{3}+j\right)}} \left._3F_2\right.
    \left(\begin{array}{ccc}
      -i & -j & \frac{5}{3}\\
      &\frac{4}{3}-i&\frac{4}{3}-j
    \end{array}\right)\,,&\nonumber
\end{eqnarray}
where $|\tilde c|^2=\sum_{i} \tilde c_i \tilde c_i$.

For $K=1$, the matrices $A,B$ become
\begin{equation}
  A_{ij}^{(K=1)}=\frac{1}{\Gamma\left(\frac{2}{3}\right)} \left(
  \begin{array}{cc}
    1 & \frac{1}{\sqrt{24}}\\
    \frac{1}{\sqrt{24}} & \frac{1}{6}
    \end{array}
  \right)\,,\quad
  B_{ij}^{(K=1)}=\frac{1}{2\Gamma\left(\frac{4}{3}\right)} \left(
  \begin{array}{cc}
    1 & \frac{1}{\sqrt{30}}\\
    \frac{1}{\sqrt{30}} & \frac{1}{3}
    \end{array}
  \right)\,,
\end{equation}
which leads to the energy eigenvalues
\begin{eqnarray}
  E_0^{(K=1)}=\frac{12}{\left(7+\sqrt{31}\right)}\times \frac{3^{\frac{1}{3}}\Gamma\left(\frac{2}{3}\right)}{\Gamma\left(\frac{1}{3}\right)}\simeq 0.696\,,\quad
  E_2^{(K=1)}=\frac{12}{\left(7-\sqrt{31}\right)}\times \frac{3^{\frac{1}{3}}\Gamma\left(\frac{2}{3}\right)}{\Gamma\left(\frac{1}{3}\right)}\simeq 6.1\,,\nonumber\\
  E_1^{(K=1)}=\frac{60}{\left(20+\sqrt{130}\right)}\times 3^{\frac{1}{3}}\Gamma\left(\frac{4}{3}\right)\simeq 2.46\,,\quad
  E_3^{(K=1)}=\frac{60}{\left(20-\sqrt{130}\right)}\times 3^{\frac{1}{3}}\Gamma\left(\frac{4}{3}\right)\simeq 9\,,\nonumber
\end{eqnarray}
and eigenvectors
\begin{eqnarray}
  \tilde c_i^{(0), (K=1)}\propto\left(\sqrt{31}+5,\sqrt{6}\right)\,,\quad
  \tilde c_i^{(2), (K=1)}\propto\left(\sqrt{31}-5,-\sqrt{6}\right)\,,\nonumber\\
  \tilde c_i^{(1), (K=1)}\propto\left(\sqrt{39}+\sqrt{30},3\right)\,,\quad
  \tilde c_i^{(3), (K=1)}\propto\left(\sqrt{39}-\sqrt{30},-3\right)\,.
\end{eqnarray}

Using the normalization condition (\ref{cnorm}), the non-vanishing matrix elements (\ref{matel}) are readily evaluated to be
\begin{eqnarray}
  \label{prek2matel}
  x_{01}&=&\frac{108 \sqrt{5}+45 \sqrt{26}+22 \sqrt{155}+10 \sqrt{806}}{\sqrt{10\left(62+11 \sqrt{31}\right)\left(130+11\sqrt{130}\right)}}\times \frac{3^{\frac{1}{3}}}{2 \sqrt{\Gamma\left(\frac{1}{3}\right)}}\simeq 0.81\,,\nonumber\\
  x_{03}&=&\frac{-108 \sqrt{5}+45 \sqrt{26}-22 \sqrt{155}+10 \sqrt{806}}{\sqrt{10\left(62+11 \sqrt{31}\right)\left(130-11\sqrt{130}\right)}}\times \frac{3^{\frac{1}{3}}}{2 \sqrt{\Gamma\left(\frac{1}{3}\right)}}\simeq -0.01\,,\nonumber\\
  x_{21}&=&\frac{-108 \sqrt{5}-45 \sqrt{26}+22 \sqrt{155}+10 \sqrt{806}}{\sqrt{10\left(62-11 \sqrt{31}\right)\left(130+11\sqrt{130}\right)}}\times \frac{3^{\frac{1}{3}}}{2 \sqrt{\Gamma\left(\frac{1}{3}\right)}}\simeq 0.87\,,\nonumber\\
  x_{23}&=&\frac{108 \sqrt{5}-45 \sqrt{26}-22 \sqrt{155}+10 \sqrt{806}}{\sqrt{10\left(62-11 \sqrt{31}\right)\left(130-11\sqrt{130}\right)}}\times \frac{3^{\frac{1}{3}}}{2 \sqrt{\Gamma\left(\frac{1}{3}\right)}}\simeq 1.65\,,
\end{eqnarray}
which are quoted in the main text.

    \section{Towards an analytic solution of the ground state of the quartic oscillator}
    \label{main}

  Using \cite[Eq.~16.4.3]{NIST:DLMF}, one can write
  \begin{equation}
    \label{eq:ars2}
  A_{ij}=\frac{1}{\Gamma^2\left(\frac{2}{3}\right)}\frac{\left(\frac{1}{3}-i\right)_j \left(\frac{1}{3}\right)_i}{\left(\frac{2}{3}-i\right)_j\left(\frac{2}{3}\right)_i}
  \sqrt{\frac{\Gamma\left(i+\frac{2}{3}\right)\Gamma\left(j+\frac{2}{3}\right)}{i! j! (1+3i)(1+3j)}}\,,
  \end{equation}
  such that using $i=0$ in Eq.~(\ref{evproblem}), the even energy-levels $E_{2n}$ can be written as
  \begin{equation}
    \label{gse}
  E_{2n}=\frac{3^{\frac{1}{3}} \Gamma\left(\frac{2}{3}\right)}{\sum_{j=0}^\infty \frac{\Gamma\left(j+\frac{1}{3}\right)}{j!}\frac{c_j^{(2n)}}{c_0^{(2n)}}}\,,
\end{equation}
where $c_i^{(2n)}$ are the unrescaled Laguerre coefficients, cf. Eq.~(\ref{rescaling}). The ground state Laguerre coefficients correspond to the dominant eigenvalue of the matrix $A_{ij}$, which can be obtained by iteration of an initial guess $c_{i}^{\rm iteration\ 0}=\delta_{i,0}$. It is convenient to introduce a second rescaling of the ground-state Laguerre coefficients,
\begin{equation}
  c_{i}^{(0)}=\frac{i!}{(1+3i)\left(\frac{2}{3}\right)_i} \tilde{\tilde{c}}_{i}\,,
\end{equation}
where using (\ref{evproblem}) and (\ref{eq:ars2}) the ground-state coefficients and energy can be found from the recursion relation
\begin{equation}
  \label{iterations}
  \tilde{\tilde{c}}_{i}^{\rm iteration\ q}=\frac{\left(\frac{1}{3}\right)_i}{i!}\sum_{j=0}^\infty \frac{\left(\frac{1}{3}-i\right)_j}{\left(\frac{2}{3}-i\right)_j}
  \frac{\tilde{\tilde{c}}_{i}^{\rm iteration\ q-1}}{1+3j}\,,\quad
  E_0^{\rm iteration\ q}=\frac{3^{\frac{1}{3}} \Gamma\left(\frac{2}{3}\right)}{\Gamma\left(\frac{1}{3}\right)\sum_{j=0}^\infty \frac{\left(\frac{1}{3}\right)_j}{(1+3j) \left(\frac{2}{3}\right)_j}\frac{\tilde{\tilde{c}}_j}{\tilde{\tilde{c}}_0}}\,.
\end{equation}

Specifically, the starting point is $\tilde{\tilde{c}}_{i}^{\rm iteration\ q}=\delta_{i,0}$, such that (\ref{gse}) implies
$$
E_0^{\rm iteration\ 0}=\frac{3^{\frac{1}{3}}\Gamma\left(\frac{2}{3}\right)}{\Gamma\left(\frac{1}{3}\right)}\simeq 0.729011\ldots
$$
The first non-trivial iteration gives
\begin{equation}
  \tilde{\tilde{c}}_{i}^{\rm iteration\ 1}=\frac{\left(\frac{1}{3}\right)_i}{i!}\,,\quad
    E_0^{\rm iteration\ 1}=\frac{3^{\frac{1}{3}}\Gamma\left(\frac{2}{3}\right)}{\Gamma\left(\frac{1}{3}\right)\left._3F_2\right.\left(\begin{array}{ccc}
        \frac{1}{3}&\frac{1}{3}&\frac{1}{3}\\
        & \frac{4}{3}&\frac{2}{3}
\end{array}\right)}\simeq 0.676893\ldots\,,
\end{equation}
where here and in the following $\left._pF_q\right.$ denotes a generalized hypergeometric function of argument unity.

The next iteration $\tilde{\tilde{c}}_{i}^{\rm iteration\ 2}=\frac{\left(\frac{1}{3}\right)_i}{i!}\left._3F_2\right.\left(\begin{array}{ccc}
        \frac{1}{3}&\frac{1}{3}&\frac{1}{3}-i\\
        & \frac{4}{3}&\frac{2}{3}-i
\end{array}\right)$, 
and $E_0^{\rm iteration\ 2}$ involves the non-standard sum
\begin{equation}
  \label{eqsum1}
\sum_{j=0}^\infty \frac{\left(\frac{1}{3}\right)_j^2}{j! (1+3j) \left(\frac{2}{3}\right)_j}\left._3F_2\right.\left(\begin{array}{ccc}
        \frac{1}{3}&\frac{1}{3}&\frac{1}{3}-j\\
        & \frac{4}{3}&\frac{2}{3}-j
\end{array}\right)\,.
\end{equation}
The generalized hypergeometric function appearing in this sum is a non-terminating Saalsch\"utzian, which using \cite[Eq.~4.1]{koornwinder1998} may be rewritten as
\begin{eqnarray}
  \label{nonterm}
  \left._3F_2\right.\left(\begin{array}{ccc}\frac{1}{3}&\frac{1}{3}&\frac{1}{3}-j;\\
    \frac{4}{3}&\frac{2}{3}-j;& 1\end{array}\right)&=&\frac{\Gamma\left(\frac{2}{3}\right) \Gamma\left(\frac{4}{3}\right)\Gamma^2\left(\frac{2}{3}+j\right)}{\Gamma\left(\frac{1}{3}+j\right)}\left[
    \frac{1}{j!}\right.\\
    &&\left.+\frac{\Gamma\left(\frac{2}{3}+j\right)}{\Gamma^2\left(\frac{1}{3}\right)\Gamma\left(\frac{4}{3}+j\right)\Gamma\left(\frac{5}{3}+j\right)} \left._3F_2\right.\left(\begin{array}{ccc}\frac{2}{3}&\frac{2}{3}+j&\frac{2}{3}+j;\\ \frac{4}{3}+j&\frac{5}{3}+j;&1\end{array}\right)\right]\,.\nonumber
\end{eqnarray}
Inserting the first part of this decomposition leads to a contribution to (\ref{eqsum1}) that is readily evaluated. For the second part, note that 
 \begin{equation}
    \sum_{j=0}^\infty \frac{\left(\frac{1}{3}\right)_j}{(1+3j) j!} t^j \left._1F_0\right.\left(\frac{2}{3},t\right)=\left._2F_1\right.\left(\frac{1}{3},\frac{1}{3},\frac{4}{3},t\right) (1-t)^{-\frac{2}{3}}=\left._2F_1\right.\left(1,1,\frac{4}{3},t\right)\,.
 \end{equation}
 Multiplying by $t^{-\frac{1}{3}}$ and integrating gives
   \begin{equation}
    \sum_{j=0}^\infty \frac{\left(\frac{1}{3}\right)_j}{(1+3j) j!} t^j \frac{\left(\frac{2}{3}\right)_j}{\left(\frac{5}{3}\right)_j}\left._2F_1\right.\left(\frac{2}{3},\frac{2}{3}+j;\frac{5}{3}+j;t\right)=\left._3F_2\right.\left(\frac{2}{3},1,1;\frac{4}{3},\frac{5}{3};t\right)\,.
   \end{equation}
   Multiplying by $t^{-\frac{1}{3}}(1-t)^{-\frac{1}{3}}$ and integrating t from $0$ to $1$ leads to
   \begin{equation}
     \label{eq:from}
    \sum_{j=0}^\infty \frac{\left(\frac{1}{3}\right)_j}{(1+3j) j!} \frac{\left(\frac{2}{3}\right)^2_j}{\left(\frac{4}{3}\right)_j\left(\frac{5}{3}\right)_j}\left._3F_2\right.\left(\frac{2}{3},\frac{2}{3}+j,\frac{2}{3}+j;\frac{4}{3}+j,\frac{5}{3}+j;1\right)=\left._4F_3\right.\left(\frac{2}{3},\frac{2}{3},1,1;\frac{4}{3},\frac{4}{3},\frac{5}{3};1\right)\,,
    \end{equation}
    such that, reassembling all parts of (\ref{eqsum1}) one finds for the second iteration ground state energy
    $$
    E_0^{\rm iteration\ 2}=\frac{3^{\frac{1}{3}}\Gamma\left(\frac{2}{3}\right)\left[1+\frac{9}{2\Gamma^3\left(\frac{1}{3}\right)}\left._3F_2\right.\left(\frac{2}{3},\frac{2}{3},\frac{2}{3};\frac{4}{3},\frac{5}{3};1\right)\right]}{\Gamma\left(\frac{1}{3}\right)\left[\left._3F_2\right.\left(\frac{1}{3},\frac{1}{3},\frac{2}{3};1,\frac{4}{3};1\right)+\frac{9}{2 \Gamma^3\left(\frac{1}{3}\right)}\left._4F_3\right.\left(\frac{2}{3},\frac{2}{3},1,1;\frac{4}{3},\frac{4}{3},\frac{5}{3};1\right) \right]}\simeq 0.66936\ldots
    $$

    At the third iteration, plugging $\tilde{\tilde{c}}_{i}^{\rm iteration\ 2}$ with the decomposition (\ref{nonterm}) into (\ref{iterations}), the sum
    \begin{equation}
  \label{eqsum2}
  I_m=\sum_{j=0}^\infty \frac{\left(\frac{1}{3}-m\right)_j}{\left(\frac{2}{3}-m\right)_j} \frac{\left(\frac{2}{3}\right)_j^3}{(1+3j) j! \left(\frac{4}{3}\right)_j \left(\frac{5}{3}\right)_j}\left._3F_2\right.\left(\begin{array}{ccc}\frac{2}{3}&\frac{2}{3}+j&\frac{2}{3}+j\\
      & \frac{4}{3}+j & \frac{5}{3}+j\end{array}\right)\,,
    \end{equation}
with $I_0=\left._4F_3\right.\left(\frac{2}{3},\frac{2}{3},1,1;\frac{4}{3},\frac{4}{3},\frac{5}{3};1\right)$  from (\ref{eq:from})  appears. Writing
    $$
    \sum_{j=0}^\infty \frac{\left(\frac{2}{3}\right)_j \left(\frac{1}{3}-m\right)_j t^j}{(1+ 3j)j!\left(\frac{2}{3}-m\right)_j }\left._1F_0\right.\left(\frac{2}{3};;t\right) =\left._3F_2\right.\left(\begin{array}{ccc}\frac{1}{3}&\frac{2}{3}&\frac{1}{3}-m;\nonumber\\
      \frac{4}{3}&\frac{2}{3}-m;&t\end{array}\right)\left._1F_0\right.\left(\frac{2}{3};;t\right) \equiv s_m(t)\,,
      $$
      the sought-after sum (\ref{eqsum2}) arises as from considering $\int_0^1 \frac{du}{u}(1-u)^{-\frac{1}{3}}\int_0^u dt t^{-\frac{1}{3}}s_m(t)$.

      A well-known property of generalized hypergeometric functions is the relations of shifting upper and lower parameters by unity. For the case at hand, it is particularly convenient to consider  $F_n(t)\equiv \left._{p+2}F_{q+2}\right.\left(\begin{array}{ccc}
    a & (b)_p & d-n\\
  1+a & (c)_q & g-n
      \end{array}\right)(t)$ which upon shifting gives the relation
      \begin{equation}
        \label{id1}
  F_n=\frac{(d-n)(g-n-a)}{(g-n)(d-n-a)} F_{n-1}
  +\frac{a (d-g)}{(g-n)(d-n-a)} \left._{p+1}F_{q+1}\right.\left(\begin{array}{cc}
    (b)_p & d-n\\
   (c)_q & g-n+1
  \end{array}\right)\,,
      \end{equation}
      where $p,q,n$ are integers and all the parameters $a,(b)_1,(b)_2,\ldots (b)_p,(c)_1,(c)_2,\ldots (c)_q,d,g$ are arbitrary. Note that (\ref{id1}) relates generalized hypergeometric functions of arbitrary argument $t$ (not written), and thus directly is applicable to $s_m(t)$ above for $a=\frac{1}{3},b=\frac{2}{3},d=\frac{1}{3},g=\frac{2}{3}$, finding the recursion relation
      $$
      s_m(t)=\frac{\left(m-\frac{1}{3}\right)^2}{m \left(m-\frac{2}{3}\right)} s_{m-1}(t)-\frac{1}{9 m \left(m-\frac{2}{3}\right)}\left._2F_1\right.\left(\begin{array}{cc} \frac{4}{3}&1-m\\
        & \frac{5}{3}-m
        \end{array}\right)(t)\,,
      $$
      where I used $\left._1F_0\right.\left(\frac{2}{3};;t\right)\left._2F_1\right.\left(\begin{array}{cc} \frac{2}{3}&\frac{1}{3}-m\\
        & \frac{5}{3}-m
        \end{array}\right)(t)=\left._2F_1\right.\left(\begin{array}{cc} \frac{4}{3}&1-m\\
        & \frac{5}{3}-m
      \end{array}\right)(t)$ to combine the last expression. Integrating twice then leads to the following recursion relation for the sum (\ref{eqsum2}):
      \begin{equation}
        \label{rec2}
        I_m=\frac{\left(m-\frac{1}{3}\right)^2}{m \left(m-\frac{2}{3}\right)} I_{m-1}-\frac{2}{27 m}\frac{\Gamma^2(m) }{\left(\frac{2}{3}\right)_m\left(\frac{1}{3}\right)_m}\,,
        \end{equation}
      where I used the Saalsch\"utzian identity \cite[Eq.~16.4.3]{NIST:DLMF} to write
      $$
      \left._3F_2\right.\left(\begin{array}{ccc} \frac{2}{3}&\frac{2}{3}&1-m\\
          & \frac{5}{3} & \frac{5}{3}-m
      \end{array}\right)=\frac{\Gamma^2(m) \Gamma\left(\frac{5}{3}\right)\Gamma\left(\frac{1}{3}\right)}{\Gamma\left(\frac{2}{3}+m\right)\Gamma\left(-\frac{2}{3}+m\right)}\,.
      $$
      Eq.~(\ref{rec2}) is a first-order recursion relation that is straightforward to solve so that (\ref{eqsum2}) is
      \begin{eqnarray}
        I_m&=&\frac{\left(\frac{2}{3}\right)_m^2}{m! \left(\frac{1}{3}\right)_m}
        \left[\left._4F_3\right.\left(\begin{array}{cccc}
            \frac{2}{3}&\frac{2}{3}&1 & 1\\
            & \frac{4}{3} & \frac{4}{3}& \frac{5}{3}
            \end{array}\right)-\frac{1}{4}\left._4F_3\right.\left(\begin{array}{cccc}
            1&1&1 & 1\\
            & \frac{5}{3} & \frac{5}{3}& \frac{5}{3}
          \end{array}\right)\right.\nonumber\\
          &&\left.
          +\frac{1}{4}\frac{m!^3}{\left(\frac{5}{3}\right)_m^3}\left._4F_3\right.\left(\begin{array}{cccc}
            1&1+m&1+m & 1+m\\
            & \frac{5}{3}+m & \frac{5}{3}+m& \frac{5}{3}+m
            \end{array}\right)
          \right]\,.
        \end{eqnarray}
      Now using relations between generalized hypergeometric functions of the form\footnote{These relations may be derived by using (\ref{id1}) with $a=\frac{1}{3},b_1=b_2=\frac{2}{3},c_1=1$ and $a=\frac{1}{3},b_1=b_2=\frac{2}{3},c_1=1,d=\frac{1}{3}-m$, respectively, for which the recursion relation for $F_n$ can be solved explicitly using the known form of the non-terminating Saalsch\"utzian \cite[Eq.~4.1]{koornwinder1998}.}
      \begin{eqnarray}
        \left._4F_3\right.\left(\begin{array}{cccc}
      \frac{2}{3}&\frac{2}{3}& 1& 1;\\
      \frac{4}{3}&\frac{4}{3}&\frac{5}{3};1
\end{array}\right)
     -\frac{1}{4}\left._4F_3\right.\left(\begin{array}{cccc}
      1&1& 1& 1;\\
      \frac{5}{3}&\frac{5}{3}&\frac{5}{3};1
     \end{array}\right)=\frac{2 \Gamma^3\left(\frac{1}{3}\right)}{9}\left._3F_2\right.\left(\begin{array}{ccc}
       \frac{1}{3}&\frac{1}{3}&\frac{2}{3}\\
       1 & \frac{4}{3}\end{array}\right)-\frac{\Gamma^5\left(\frac{1}{3}\right)}{27 \Gamma\left(\frac{2}{3}\right)}\,,\qquad\quad
     \nonumber
      \end{eqnarray}
      \begin{eqnarray}
     \left._4F_3\right.\left(\begin{array}{cccc}
      1 &1\!+m& 1\!+m& 1\!+m;\\
     & \frac{5}{3}+m&\frac{5}{3}+m&\frac{5}{3}+m 
     \end{array}\right)=\nonumber\\
     -\frac{3 \left(\frac{2}{3}+m\right)^2\Gamma^2\left(\frac{1}{3}\right)\Gamma\left(\frac{1}{3}+m\right)\Gamma\left(\frac{5}{3}+m\right)}{\Gamma\left(\frac{2}{3}\right)m!^2}
    \left._4F_3\right.\left(\begin{array}{cccc}
      \frac{1}{3} &\frac{2}{3}&\frac{2}{3}&\frac{1}{3}-m\\
      &1 & \frac{4}{3} & \frac{2}{3}-m
    \end{array}\right)\nonumber\\
    +\frac{\Gamma^2\left(\frac{1}{3}\right)\Gamma^2\left(\frac{5}{3}+m\right)}{\Gamma^2\left(\frac{2}{3}\right)\Gamma^2\left(\frac{4}{3}+m\right)}\left._4F_3\right.\left(\begin{array}{cccc}
      \frac{2}{3} &\frac{2}{3}+m&1+m&1+m\\
      &\frac{4}{3}+m & \frac{4}{3}+m & \frac{5}{3}+m
    \end{array}\right)
    +\frac{\Gamma^5\left(\frac{1}{3}\right)\Gamma^3\left(\frac{5}{3}+m\right)}{2 m!^3 \Gamma^4\left(\frac{2}{3}\right)}\,,\qquad\qquad
      \end{eqnarray}
      the whole third-order iteration for the Laguerre coefficients can be written as
      $$
\tilde{\tilde{c}}_i^{\rm iteration\ 3}=\frac{\left(\frac{2}{3}\right)_i}{i!}\left[\frac{\left(\frac{2}{3}\right)_i}{i!}
  \left._3F_2\right.\left(\begin{array}{ccc}
    \frac{1}{3} & \frac{1}{3} & \frac{2}{3}\\
    & 1 & \frac{4}{3}
  \end{array}\right)
  +\frac{9}{2 \Gamma^3\left(\frac{1}{3}\right)} \frac{(1)_i^2 \left(\frac{2}{3}\right)_i}
  {\left(\frac{4}{3}\right)_i^2 \left(\frac{5}{3}\right)_i}
  \left._4F_3\right.\left(\begin{array}{cccc}
    \frac{2}{3} & 1+i & 1+i & \frac{2}{3}+i\\
    & \frac{4}{3}+i & \frac{4}{3}+i& \frac{5}{3}+i
    \end{array}\right)
  \right]\,.
$$
The third iteration of the ground state energy involves again a sum that can be solved with the same methods as those outlined above and one finds
\begin{equation}
  E_0^{\rm iteration\ 3}= \frac{3^{1/3}\Gamma\left(\frac{2}{3}\right)}{\Gamma\left(\frac{1}{3}\right)}
\frac{\left._3F_2\right.\left(\begin{array}{ccc}
       \frac{1}{3}&\frac{1}{3}&\frac{2}{3}\\
       & 1 & \frac{4}{3}
\end{array}\right)+\frac{9}{2\Gamma^3\left(\frac{1}{3}\right)}\left._4F_3\right.\left(\begin{array}{cccc}
       \frac{2}{3}&\frac{2}{3}&1 & 1\\
       & \frac{4}{3} & \frac{4}{3} & \frac{5}{3}
  \end{array}\right)}
  {\left(\left._3F_2\right.\left(\begin{array}{ccc}
       \frac{1}{3}&\frac{1}{3}&\frac{2}{3}\\
       & 1 & \frac{4}{3}
\end{array}\right)\right)^2+\frac{9}{2\Gamma^3\left(\frac{1}{3}\right)}\left._5F_4\right.\left(\begin{array}{ccccc}
  \frac{2}{3} & 1 & 1 & 1 &1\\
  & \frac{4}{3} & \frac{4}{3} & \frac{4}{3} & \frac{5}{3}
  \end{array}\right)}\simeq 0.6683\,.
  \end{equation}

The fourth iteration again leads to a recursion relation similar to that for (\ref{eqsum1}), but now involving $\left._4F_3\right.\left(\begin{array}{cccc}     \frac{2}{3} & 1 & 1  & -j\\ 
     & \frac{4}{3} & \frac{5}{3}  & \frac{2}{3}-j
  \end{array}\right)$. Unlike the case for the non-terminating Saalsch\"utzian $\left._3F_2\right.$, there is no known general formula to expand the non-terminating Saalsch\"utzian $\left._4F_3\right.$, so I did not find a closed-form expression for $\tilde{\tilde{c}}_i^{\rm iteration\ 4}$, despite its close similarity with the sum appearing in $E_0^{\rm iteration\ 3}$.

\section{Exact high temperature limit}
\label{high}

Recall that the partition function for quantum mechanics with Hamiltonian (\ref{hamiltonian}) can be written in terms of a path integral \cite{Laine:2016hma}[Eq.~1.35],
\begin{equation}
  Z(\beta)=\lim_{N\rightarrow \infty}\int \left[\prod_{i=1}^N \frac{d x_i}{\sqrt{4 \pi \beta/N}}\right] e^{-\frac{\beta}{N}\sum_{j=1}^N\left[\frac{1}{4}\left(\frac{x_{j+1}-x_j}{\beta/N}\right)^2+V(x_j)\right]}\,,
\end{equation}
with periodic boundary conditions $x_{N+1}=x_1$. In the high temperature limit $\beta\rightarrow 0$, only the Fourier zero-mode $x_1=x_2=\ldots=\bar x={\rm const}.$ contributes to $Z(\beta)$. This is the same as having only one site, $N=1$
As a consequence, one has
\begin{equation}
  \lim_{\beta\rightarrow 0}Z(\beta)=\int_{-\infty}^\infty \frac{d\bar x}{\sqrt{4 \pi \beta}} e^{-\beta V(\bar x)}\,,
  \end{equation}
which can be evaluated for any potential $V(x)$. Specifically, for $V(x)=\frac{x^4}{4}$ one finds $Z(\beta)=\beta^{-\frac{3}{4}}\times {\rm const}$, such that the spectral form factor (\ref{ssf}) in the high temperature limit becomes
\begin{equation}
  \lim_{\beta\rightarrow 0} g(\beta,t)=\left(1+\beta^{-2}t^2\right)^{-\frac{3}{4}}\,.
\label{bsf}
  \end{equation}

  \end{appendix}

\bibliography{otoc}

\begin{thebibliography}{12}%
\makeatletter
\providecommand \@ifxundefined [1]{%
 \@ifx{#1\undefined}
}%
\providecommand \@ifnum [1]{%
 \ifnum #1\expandafter \@firstoftwo
 \else \expandafter \@secondoftwo
 \fi
}%
\providecommand \@ifx [1]{%
 \ifx #1\expandafter \@firstoftwo
 \else \expandafter \@secondoftwo
 \fi
}%
\providecommand \natexlab [1]{#1}%
\providecommand \enquote  [1]{``#1''}%
\providecommand \bibnamefont  [1]{#1}%
\providecommand \bibfnamefont [1]{#1}%
\providecommand \citenamefont [1]{#1}%
\providecommand \href@noop [0]{\@secondoftwo}%
\providecommand \href [0]{\begingroup \@sanitize@url \@href}%
\providecommand \@href[1]{\@@startlink{#1}\@@href}%
\providecommand \@@href[1]{\endgroup#1\@@endlink}%
\providecommand \@sanitize@url [0]{\catcode `\\12\catcode `\$12\catcode
  `\&12\catcode `\#12\catcode `\^12\catcode `\_12\catcode `\%12\relax}%
\providecommand \@@startlink[1]{}%
\providecommand \@@endlink[0]{}%
\providecommand \url  [0]{\begingroup\@sanitize@url \@url }%
\providecommand \@url [1]{\endgroup\@href {#1}{\urlprefix }}%
\providecommand \urlprefix  [0]{URL }%
\providecommand \Eprint [0]{\href }%
\providecommand \doibase [0]{http://dx.doi.org/}%
\providecommand \selectlanguage [0]{\@gobble}%
\providecommand \bibinfo  [0]{\@secondoftwo}%
\providecommand \bibfield  [0]{\@secondoftwo}%
\providecommand \translation [1]{[#1]}%
\providecommand \BibitemOpen [0]{}%
\providecommand \bibitemStop [0]{}%
\providecommand \bibitemNoStop [0]{.\EOS\space}%
\providecommand \EOS [0]{\spacefactor3000\relax}%
\providecommand \BibitemShut  [1]{\csname bibitem#1\endcsname}%
\let\auto@bib@innerbib\@empty
\bibitem [{\citenamefont {Hashimoto}\ \emph {et~al.}(2017)\citenamefont
  {Hashimoto}, \citenamefont {Murata},\ and\ \citenamefont
  {Yoshii}}]{Hashimoto:2017oit}%
  \BibitemOpen
  \bibfield  {author} {\bibinfo {author} {\bibfnamefont {Koji}\ \bibnamefont
  {Hashimoto}}, \bibinfo {author} {\bibfnamefont {Keiju}\ \bibnamefont
  {Murata}}, \ and\ \bibinfo {author} {\bibfnamefont {Ryosuke}\ \bibnamefont
  {Yoshii}},\ }\bibfield  {title} {\enquote {\bibinfo {title}
  {{Out-of-time-order correlators in quantum mechanics}},}\ }\href {\doibase
  10.1007/JHEP10(2017)138} {\bibfield  {journal} {\bibinfo  {journal} {JHEP}\
  }\textbf {\bibinfo {volume} {10}},\ \bibinfo {pages} {138} (\bibinfo {year}
  {2017})},\ \Eprint {http://arxiv.org/abs/1703.09435} {arXiv:1703.09435
  [hep-th]} \BibitemShut {NoStop}%
\bibitem [{\citenamefont {Akutagawa}\ \emph {et~al.}(2020)\citenamefont
  {Akutagawa}, \citenamefont {Hashimoto}, \citenamefont {Sasaki},\ and\
  \citenamefont {Watanabe}}]{Akutagawa:2020qbj}%
  \BibitemOpen
  \bibfield  {author} {\bibinfo {author} {\bibfnamefont {Tetsuya}\ \bibnamefont
  {Akutagawa}}, \bibinfo {author} {\bibfnamefont {Koji}\ \bibnamefont
  {Hashimoto}}, \bibinfo {author} {\bibfnamefont {Toshiaki}\ \bibnamefont
  {Sasaki}}, \ and\ \bibinfo {author} {\bibfnamefont {Ryota}\ \bibnamefont
  {Watanabe}},\ }\bibfield  {title} {\enquote {\bibinfo {title}
  {{Out-of-time-order correlator in coupled harmonic oscillators}},}\
  }\href@noop {} {\  (\bibinfo {year} {2020})},\ \Eprint
  {http://arxiv.org/abs/2004.04381} {arXiv:2004.04381 [hep-th]} \BibitemShut
  {NoStop}%
\bibitem [{\citenamefont {Bhattacharyya}\ \emph {et~al.}(2020)\citenamefont
  {Bhattacharyya}, \citenamefont {Chemissany}, \citenamefont {Haque},
  \citenamefont {Murugan},\ and\ \citenamefont {Yan}}]{Bhattacharyya:2020art}%
  \BibitemOpen
  \bibfield  {author} {\bibinfo {author} {\bibfnamefont {Arpan}\ \bibnamefont
  {Bhattacharyya}}, \bibinfo {author} {\bibfnamefont {Wissam}\ \bibnamefont
  {Chemissany}}, \bibinfo {author} {\bibfnamefont {S.~Shajidul}\ \bibnamefont
  {Haque}}, \bibinfo {author} {\bibfnamefont {Jeff}\ \bibnamefont {Murugan}}, \
  and\ \bibinfo {author} {\bibfnamefont {Bin}\ \bibnamefont {Yan}},\ }\bibfield
   {title} {\enquote {\bibinfo {title} {{The Multi-faceted Inverted Harmonic
  Oscillator: Chaos and Complexity}},}\ }\href@noop {} {\  (\bibinfo {year}
  {2020})},\ \Eprint {http://arxiv.org/abs/2007.01232} {arXiv:2007.01232
  [hep-th]} \BibitemShut {NoStop}%
\bibitem [{\citenamefont {Hashimoto}\ \emph {et~al.}(2020)\citenamefont
  {Hashimoto}, \citenamefont {Huh}, \citenamefont {Kim},\ and\ \citenamefont
  {Watanabe}}]{Hashimoto:2020xfr}%
  \BibitemOpen
  \bibfield  {author} {\bibinfo {author} {\bibfnamefont {Koji}\ \bibnamefont
  {Hashimoto}}, \bibinfo {author} {\bibfnamefont {Kyoung-Bum}\ \bibnamefont
  {Huh}}, \bibinfo {author} {\bibfnamefont {Keun-Young}\ \bibnamefont {Kim}}, \
  and\ \bibinfo {author} {\bibfnamefont {Ryota}\ \bibnamefont {Watanabe}},\
  }\bibfield  {title} {\enquote {\bibinfo {title} {{Exponential growth of
  out-of-time-order correlator without chaos: inverted harmonic oscillator}},}\
  }\href@noop {} {\  (\bibinfo {year} {2020})},\ \Eprint
  {http://arxiv.org/abs/2007.04746} {arXiv:2007.04746 [hep-th]} \BibitemShut
  {NoStop}%
\bibitem [{\citenamefont {Dyer}\ and\ \citenamefont
  {Gur-Ari}(2017)}]{Dyer:2016pou}%
  \BibitemOpen
  \bibfield  {author} {\bibinfo {author} {\bibfnamefont {Ethan}\ \bibnamefont
  {Dyer}}\ and\ \bibinfo {author} {\bibfnamefont {Guy}\ \bibnamefont
  {Gur-Ari}},\ }\bibfield  {title} {\enquote {\bibinfo {title} {{2D CFT
  Partition Functions at Late Times}},}\ }\href {\doibase
  10.1007/JHEP08(2017)075} {\bibfield  {journal} {\bibinfo  {journal} {JHEP}\
  }\textbf {\bibinfo {volume} {08}},\ \bibinfo {pages} {075} (\bibinfo {year}
  {2017})},\ \Eprint {http://arxiv.org/abs/1611.04592} {arXiv:1611.04592
  [hep-th]} \BibitemShut {NoStop}%
\bibitem [{\citenamefont {Carlitz}(1961)}]{Carlitz}%
  \BibitemOpen
  \bibfield  {author} {\bibinfo {author} {\bibfnamefont {L.}~\bibnamefont
  {Carlitz}},\ }\bibfield  {title} {\enquote {\bibinfo {title} {{Some integrals
  containing products of legendre polynomials}},}\ }\href {\doibase
  https://doi.org/10.1007/BF01650571} {\bibfield  {journal} {\bibinfo
  {journal} {Arch. Math}\ }\textbf {\bibinfo {volume} {12}},\ \bibinfo {pages}
  {334--340} (\bibinfo {year} {1961})}\BibitemShut {NoStop}%
\bibitem [{{\relax DLMF}()}]{NIST:DLMF}%
  \BibitemOpen
  {\relax DLMF},\ \href {http://dlmf.nist.gov/31.12} {\enquote {\bibinfo
  {title} {{\it NIST Digital Library of Mathematical Functions}},}\ }\bibinfo
  {howpublished} {http://dlmf.nist.gov/, Release 1.0.25 of 2019-12-15},\
  \bibinfo {note} {f.~W.~J. Olver, A.~B. {Olde Daalhuis}, D.~W. Lozier, B.~I.
  Schneider, R.~F. Boisvert, C.~W. Clark, B.~R. Miller, B.~V. Saunders, H.~S.
  Cohl, and M.~A. McClain, eds.}\BibitemShut {Stop}%
\bibitem [{\citenamefont {Bender}\ \emph {et~al.}(1977)\citenamefont {Bender},
  \citenamefont {Olaussen},\ and\ \citenamefont {Wang}}]{Bender:1977dr}%
  \BibitemOpen
  \bibfield  {author} {\bibinfo {author} {\bibfnamefont {Carl~M.}\ \bibnamefont
  {Bender}}, \bibinfo {author} {\bibfnamefont {K.}~\bibnamefont {Olaussen}}, \
  and\ \bibinfo {author} {\bibfnamefont {P.~S.}\ \bibnamefont {Wang}},\
  }\bibfield  {title} {\enquote {\bibinfo {title} {{Numerological Analysis of
  the WKB Approximation in Large Order}},}\ }\href {\doibase
  10.1103/PhysRevD.16.1740} {\bibfield  {journal} {\bibinfo  {journal} {Phys.
  Rev.}\ }\textbf {\bibinfo {volume} {D16}},\ \bibinfo {pages} {1740--1748}
  (\bibinfo {year} {1977})}\BibitemShut {NoStop}%
\bibitem [{\citenamefont {Hioe}\ \emph {et~al.}(1978)\citenamefont {Hioe},
  \citenamefont {Macmillen},\ and\ \citenamefont {Montroll}}]{Hioe:1978jj}%
  \BibitemOpen
  \bibfield  {author} {\bibinfo {author} {\bibfnamefont {F.~T.}\ \bibnamefont
  {Hioe}}, \bibinfo {author} {\bibfnamefont {D.}~\bibnamefont {Macmillen}}, \
  and\ \bibinfo {author} {\bibfnamefont {E.~W.}\ \bibnamefont {Montroll}},\
  }\bibfield  {title} {\enquote {\bibinfo {title} {{Quantum Theory of
  Anharmonic Oscillators: Energy Levels of a Single and a Pair of Coupled
  Oscillators with Quartic Coupling}},}\ }\href {\doibase
  10.1016/0370-1573(78)90097-2} {\bibfield  {journal} {\bibinfo  {journal}
  {Phys. Rept.}\ }\textbf {\bibinfo {volume} {43}},\ \bibinfo {pages}
  {305--335} (\bibinfo {year} {1978})}\BibitemShut {NoStop}%
\bibitem [{\citenamefont {Romatschke}()}]{codedown}%
  \BibitemOpen
  \bibfield  {author} {\bibinfo {author} {\bibfnamefont {P.}~\bibnamefont
  {Romatschke}},\ }\bibfield  {title} {\enquote {\bibinfo {title} {{OTOC solver
  for the quartic quantum oscillator}},}\ }\href
  {https://github.com/paro8929/OTOC} {\bibinfo  {journal}
  {https://github.com/paro8929/OTOC$\quad$}\ }\BibitemShut {NoStop}%
\bibitem [{\citenamefont {{Koornwinder}}(1998)}]{koornwinder1998}%
  \BibitemOpen
\bibfield  {journal} {  }\bibfield  {author} {\bibinfo {author} {\bibfnamefont
  {Tom~H.}\ \bibnamefont {{Koornwinder}}},\ }\bibfield  {title} {\enquote
  {\bibinfo {title} {{Identities of nonterminating series by Zeilberger's
  algorithm}},}\ }\href@noop {} {\bibfield  {journal} {\bibinfo  {journal}
  {arXiv Mathematics e-prints}\ ,\ \bibinfo {eid} {math/9805010}} (\bibinfo
  {year} {1998})},\ \Eprint {http://arxiv.org/abs/math/9805010}
  {arXiv:math/9805010 [math.CA]} \BibitemShut {NoStop}%
\bibitem [{\citenamefont {Laine}\ and\ \citenamefont
  {Vuorinen}(2016)}]{Laine:2016hma}%
  \BibitemOpen
  \bibfield  {author} {\bibinfo {author} {\bibfnamefont {Mikko}\ \bibnamefont
  {Laine}}\ and\ \bibinfo {author} {\bibfnamefont {Aleksi}\ \bibnamefont
  {Vuorinen}},\ }\href {\doibase 10.1007/978-3-319-31933-9} {\emph {\bibinfo
  {title} {{Basics of Thermal Field Theory}}}},\ Vol.\ \bibinfo {volume} {925}\
  (\bibinfo  {publisher} {Springer},\ \bibinfo {year} {2016})\ \Eprint
  {http://arxiv.org/abs/1701.01554} {arXiv:1701.01554 [hep-ph]} \BibitemShut
  {NoStop}%
\end{thebibliography}%
\end{document}